\documentclass[a4paper,fleqn]{cas-dc}

\usepackage{ulem}
\usepackage{xcolor}
\usepackage{graphicx,caption,subcaption}
\usepackage{bm}
\usepackage{microtype}

\usepackage{subcaption}
\usepackage[utf8]{inputenc}
\usepackage[english]{babel}
\usepackage{comment}
\usepackage{multirow}
\usepackage[none]{hyphenat}
\usepackage{indentfirst}

\renewcommand{\vec}[1]{\ensuremath{#1}}
\graphicspath{ {./figures/} }
\DeclareGraphicsExtensions{.png}
\usepackage{lipsum}
\usepackage[numbers]{natbib}
\renewcommand{\vec}[1]{\ensuremath{#1}}
\usepackage{dutchcal}
\usepackage{listings}
\usepackage{scalerel}
\usepackage[ruled,longend]{algorithm2e}
\usepackage{amsmath}
\usepackage{scalerel}

\newcommand\reallywidehat[1]{\arraycolsep=0pt\relax%
\begin{array}{c}
\stretchto{
  \scaleto{
    \scalerel*[\widthof{\ensuremath{#1}}]{\kern-.5pt\bigwedge\kern-.5pt}
    {\rule[-\textheight/2]{1ex}{\textheight}} %WIDTH-LIMITED BIG WEDGE
  }{\textheight} % 
}{0.5ex}\\           % THIS SQUEEZES THE WEDGE TO 0.5ex HEIGHT
#1\\                 % THIS STACKS THE WEDGE ATOP THE ARGUMENT
\rule{-1ex}{0ex}
\end{array}
}
\usepackage[numbers]{natbib}

\def\tsc#1{\csdef{#1}{\textsc{\lowercase{#1}}\xspace}}
\tsc{WGM}
\tsc{QE}
\tsc{EP}
\tsc{PMS}
\tsc{BEC}
\tsc{DE}
\begin{document}
\begin{sloppypar}

\let\WriteBookmarks\relax
\def\floatpagepagefraction{1}
\def\textpagefraction{.001}
% \shorttitle{Composite Structures}
\shorttitle{Minimal mass design of a tensegrity tower for lunar electromagnetic launching}
\shortauthors{X. Su, M. Chen, M. Majji and R. E. Skelton}

\title [mode = title]{Minimal mass design of a tensegrity tower for lunar electromagnetic launching}
% \title [mode = title]{Design and analysis of an electromagnetic tensegrity lunar launcher}

\author[1]{Xiaowen Su}[type=editor,
auid=000,bioid=1,
orcid=0000-0001-6922-4748]
\fnmark[1]
%\ead{suxiaowen0508@gmail.com}

\fntext[1]{Postdoctoral Researcher, Department of Aerospace Engineering, Texas A\&M University, College Station, TX, USA.}

\author[1]{Muhao Chen}[type=editor,
auid=000,bioid=1,
orcid=0000-0003-1812-6835]
\fnmark[1]
\cormark[1]
\ead{muhaochen@tamu.edu}

\address[1]{Department of Aerospace Engineering, Texas A\&M University, College Station, TX, USA}

\cortext[cor1]{Corresponding author. Tel.: +1 9799858285.}

% \fntext[fn2]{Postdoctoral Researcher, Department of Aerospace Engineering, Texas A\&M University, College Station, TX, USA.}

\author[1]{Manoranjan Majji}[orcid=0000-0002-8425-8687]
\fnmark[3]
%\ead{mmajji@tamu.edu}

\fntext[fn3]{Director of LASR Lab, Assistant Professor, Department of Aerospace Engineering, Texas A\&M University, College Station, TX, USA.}

\author[1]{Robert E. Skelton}[orcid=0000-0001-6503-9115]
\fnmark[4]
% \ead{bobskelton@tamu.edu}

\fntext[fn4]{TEES Eminent Professor, Department of Aerospace Engineering, Texas A\&M University, College Station, TX, USA.}

\begin{abstract}
\doublespacing
Lunar explorations have provided us with information about its abundant resources that can be utilized in orbiting-resource depots as lunar-derived commodities. To reduce the energy requirements of a launcher to send these commodities from the lunar surface to the space depots, this paper explores the application of the electromagnetic acceleration principle and provides an assessment of the actual technical characteristics of the launcher's installation to ensure the acceleration of a payload with a mass of 1,500 kg to a speed of 2,200 m/s (circumlunar orbit speed). To fulfill a lightweight (fewer materials and less energy) support structure for the electromagnetic launcher with strength requirements, the tensegrity structure minimum mass principle without global buckling has been developed and applied to support the electromagnetic acceleration device. Therefore, this paper proposes and develops a minimal mass electromagnetic tensegrity lunar launcher. We first demonstrate the mechanics of launcher and payload, how a payload can be accelerated to a specific velocity, and how a payload carrier can be recycled for another launch. Then, a detailed discussion on the lunar launch system, procedures of propulsion, the required mass, and energy of the launch barrel are given. The governing equations of tensegrity minimal mass tensegrity design algorithm with gravity and without global buckling. Finally, a case study is conducted to show a feasible structure design, the required mass, and energy. The principles developed in this paper are also applicable to the rocket launch system, space elevator, space train transportation, interstellar payload package delivery, etc.
\end{abstract}


\maketitle

\section{Introduction}

Lunar explorations have provided us with information about its abundant useful resources in the lunar soils and its polar craters. For example, the lunar soils (such as regolith, aluminum, chromium, nitrogen, and oxygen) can be utilized to produce rare metals and fuels, and amounts of ice in the lunar polar craters can produce water \cite{hoshino2020lunar,Moon2014}. There is a high price to pay in launching the commodities from Earth via liquid propellants than from the moon via solar energy to the depots to facilitate some deep space missions, such as space station construction and interplanetary travel. The first price is that Earth's gravity (lunar gravity is 1/6 of earth gravity) and atmosphere (nearly no atmosphere on the moon) cause more energy losses in getting a payload mass to orbit. The second one is that fuels are valuable commodities, and there is no reason to consume valuable fuel instead of using more available renewable solar energy. In fact, many lunar-resource-utilization researches have been conducted, such as rover for lunar landing \cite{sutoh2017rover}, robotic operations on the moon \cite{austin2020robotic}, lunar base design principles \cite{sherwood2019principles}, 3D printing by regolith \cite{meurisse2018solar}, energy storage and generation on the moon \cite{palos2020lunar}, etc. Furthermore, researchers have proposed to use the lunar-derived commodities instead of earth-commodities in the orbiting-resource depots to enable human space dwelling, space plant cultivation, essential life supplements, and refuel landers or other interplanetary vehicles in the future \cite{utilization2011}. However, how to deliver the lunar-derived commodities to the orbiting-resource depots still needs to be explored.\par 

\textbf{EML:} The electromagnetic lunar launcher technology can be the answer to this dilemma. At first, the electromagnetic acceleration principle can use renewable solar-electrical energy instead of non-renewable chemical energy. Secondly, the availability of solar energy on the lunar surface provides higher repeatability of the electromagnetic launcher rather than its counterpart, a traditional chemical energy-based launcher. Thirdly, the non-contact electromagnetic acceleration technology can allow a lower chance of maintenance comparing with a traditional rocket engine with the combustion of reactive chemicals. Finally, the further development of this electromagnetic acceleration technology in high-speed transportation sectors can be useful on many other space missions. A few kinds of research have been conducted on electromagnetic launching (EML) systems. For example, EML was first proposed to be an alternative launch medium by  \cite{foner1957coils,hawke1982electromagnetic,Henry1982,kolm1984basic}, and then the capability of utilizing an electromagnetically levitated vehicle on a horizontal linear motor track has been studied by NASA for twenty years \cite{Jon2003}. Later, the tradeoffs between the design and control of the electromagnetic coilguns (EML with rails) were discussed, resulting in a set of operational requirements \cite{Kaye2005}. Ismagilov et al. designed a high-speed permanent magnet generator with an amorphous alloy magnetic core for aerospace applications \cite{ismagilov2019design}, and presented an algorithm for the ultra-high-speed electrical machine (UHSEM) \cite{ismagilov2018multidisciplinary}. Inger evaluated the electromagnetic launching to geosynchronously equatorial orbit and its cost \cite{inger2017electromagnetic}. Engeland Prelas showed that the EML technology was sufficiently matured for asteroid mining and deflection applications \cite{engel2017asteroid}. Yang et al. designed and tested the coil-unit barrel \cite{yang2018design}, and studied the relationship among commutation-induced voltage (CIV), current, the commutation inductance gradient (CIG), and the projectile velocity for helical coil electromagnetic launchers (HEMLs) \cite{yang2016research}. Abdo et al. coupled the mechanical and magnetic finite element analysis models to examine the effects of changing the capacitor voltage level and the initial position of the projectile on the acceleration, speed, and force \cite{abdo2016performance}. In conclusion, to reduce the energy cost of a launcher to send these commodities from the lunar surface to the space depots, this paper explores the application of the electromagnetic acceleration principle and provides an assessment of the actual technical characteristics of the launcher's installation to ensure the acceleration of a payload with a mass of 1,500 kg to a speed of 2,200 m/s (circumlunar orbit speed).\par

The EML can be applied to both micro-scale and large-scale applications. For micro-scale application, an atomic coilgun was studied by \cite{narevicius2008stopping}. For a large-scale application, the EML has the potential to launch a payload to a specific exit velocity, but the launcher would take a large compressive load (propulsion force). To design a support structure, met the strength requirements for a long-barrel electromagnetic launcher, a considerable amount of supporting material is required to prevent the long launch barrel from buckling. To save material, we seek the solution of the minimum mass electromagnetic launcher. \par

\textbf{Tensegrity:} Originally, the tensegrity art-form was first created by Ioganson (1921) and Snelson (1948) \cite{Snelson_1965}, and the word \lq tensegrity\rq ~is coined from two words \lq tensile + integrity = tensegrity\rq~by Buckminster Fuller \cite{Fuller_1959}. After decades of study, tensegrity has shown its advantages in minimal mass \cite{chen2020habitat,ma2020design,wang2021minimal}, modularity, deployability \cite{ma2021design}, high precision control \cite{rieffel2018adaptive}, redundancy in actuation, abundant equilibrium states \cite{ma2021tensegrity}, achieving different stiffness by changing string prestress \cite{sultan2009tensegrity}, promoting the integration of structure and control design \cite{fraternali2014multiscale,peng2020novel,zheng2021robustness}. This paper explores a tensegrity tower to facilitate the application of electromagnetic launching on the moon. In fact, A few studies have been made on tensegrity tower designs. For example, Sultan and Skelton demonstrated a deployment strategy for tensegrity structures by a multi-stage three-strut Snelson-type tensegrity tower \cite{sultan2003deployment}. Klimke and Soeren presented a construction process of a tensegrity tower \cite{klimke2004making}. Chen et al. showed a deployable tower for taking compressive load at the top of the tower \cite{chen2020deployable}. Yildiz and Lesieutre studied the approach to obtain effective continuum beam stiffness properties of tensegrity towers with n struts \cite{yildiz2019effective}. However, none of these towers modeled gravity in the design process and viewed mass design and local and global stability as an integrated design process. In order to achieve a small amount of valuable supporting resources/mass in space as possible, this paper presents a minimal mass design algorithm subject to gravity and structure stability for any tensegrity structures. By integrating the tensegrity design algorithm and electromagnetic principles, a lightweight tensegrity lunar launcher is designed and analyzed.

The rest of this paper is organized as follows: Section 2 describes the mechanics of the launcher and payload. Section 3 presents the lunar launch system, procedures of propulsion, the required mass, and the energy of the launch barrel. Section 4 gives the tensegrity notations and minimal mass tensegrity design algorithm. Section 5 conducts a case study to determine the launcher system's mass, geometry, energy, and structure complexity, as well as the required structure mass. Section 6 summarizes the conclusions.

\section{The mechanics of launcher and payload}
% given exit velocity 
% \subsection{Design Objective}
We apply the electromagnetic propulsion principle to launch a loaded projectile, which includes a payload installed in a carrier, to a lunar escape velocity, and then to decelerate the empty projectile (only contains the payload carrier) at the top of a launch tower for recycling.
Given the payload mass $m_p$, the payload-carrier mass $m_{pc}$, the acceleration $a_1$ and the exit velocity $v_e$, we can obtain the required propulsion force $f$, the deceleration $a_2$, the total launch height $H$ and the total launch time $T$. They are calculated as:
\begin{align}\label{f}
    f & = (m_p+m_{pc})a_1,\\ \label{a2}
    a_2 & = \frac{m_p+m_{pc}}{m_{pc}}a_1,\\ \label{H}
    H & = H_a+H_d = \frac{{v_e}^2}{2a_1}+\frac{{v_e}^2}{2a_2}, \\      \label{T}
     T & = t_1+t_2 = \frac{v_e}{a_1}+\frac{v_e}{a_2},
\end{align}
where $a_1$, $H_a$ and $t_1$ are the loaded projectile acceleration, acceleration height and acceleration time, respectively. $a_2$, $H_d$ and $t_2$ are the empty projectile deceleration, deceleration height and deceleration time, respectively. $T$ and $H$ are the total time and height through launching. Note that the acceleration force and the deceleration force have the same magnitude but different directions. It will be explained in Section 3.1.

\section{Electromagnetic propulsion}
In this section, we will introduce how the electromagnetic propulsion principle can work in this case and what is the required mass and energy of the electromagnetic propulsion system in this application. 
\subsection{Launch system and procedures of propulsion}
\begin{figure}
\centering
\includegraphics[scale=0.5]{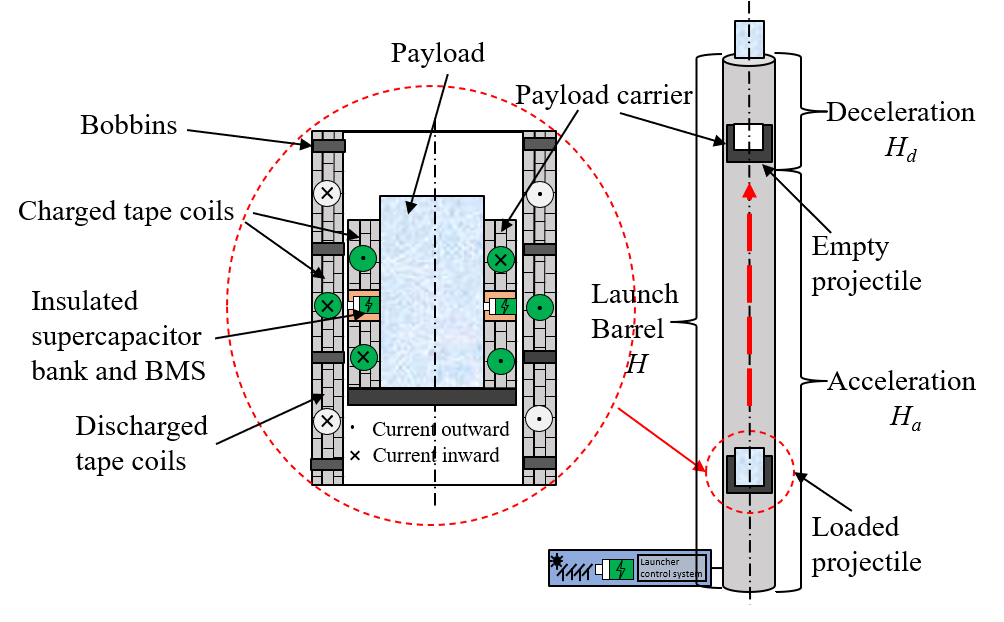}
\caption{Electromagnetic launch system, the projectile locates inside the launch barrel. Other supporting systems include solar panels, battery management systems, and control systems.}
\label{fig:procedure} 
\end{figure}

As shown in Figure \ref{fig:procedure}, the structure of the electromagnetic propulsion system is mainly composed of a launch barrel and a projectile (loaded or empty). The launch barrel is composed of some bobbins that can be stacked together. There are some coils winding around the bobbins. To generate the required amount of electromagnetic propulsion force, the coils on the launch barrel can be charged or discharged. \cite{wang2013practical}. The projectile is composed of a payload (i.e., ice, regolith, minerals), a payload carrier with upper and lower coils powered by supercapacitors, and a battery management system. There is some insulation coating for thermal and cosmic radiation protection attached to the projectile.

\par During acceleration, the loaded projectile is accelerated to the lunar escape velocity by a set of continuous electromagnetic forces generated between some charged coils on the payload carrier and those on the launch barrel. The payload leaves its carrier and keeps moving at the lunar escape velocity at height $H_a$ until it exits the launch barrel. And then, the empty projectile is decelerated by the electromagnetic force, similar to the acceleration force in magnitude and different from that in directions. The direction of the electromagnetic force can be alternated by changing the current direction in the coils on the launch barrel. And then, to reuse the payload-carrier, it can be designed to pull down from top to bottom inside the launch barrel by a mechanism; Finally, the launch system required energy will be restored in an energy storage system via some solar panels for another launch; another payload will be installed in the payload-carrier; an on-the-ground power system can charge the coils of the bobbins sequentially on/off from the bottom to the top of the launch barrel to provide a continuous electromagnetic propulsion force; the supercapacitors can be recharged to power the payload-carrier coils. \par 

It is worth mentioning that since solar panels will store energy in batteries, a suitable battery management system (BMS) is vital in ensuring batteries' safe and reliable operation. Critical technologies in the BMS include battery modeling, internal state estimation, and battery charging \cite{liu2019brief}. The battery and BMS designs are beyond this research scope, but related studies can be found in \cite{liu2017design,wang2018hybrid,xinlin2019research}.
The design of the corresponding control system is out of the scope of this paper. 
\begin{figure}
\centering
\includegraphics[scale=0.55]{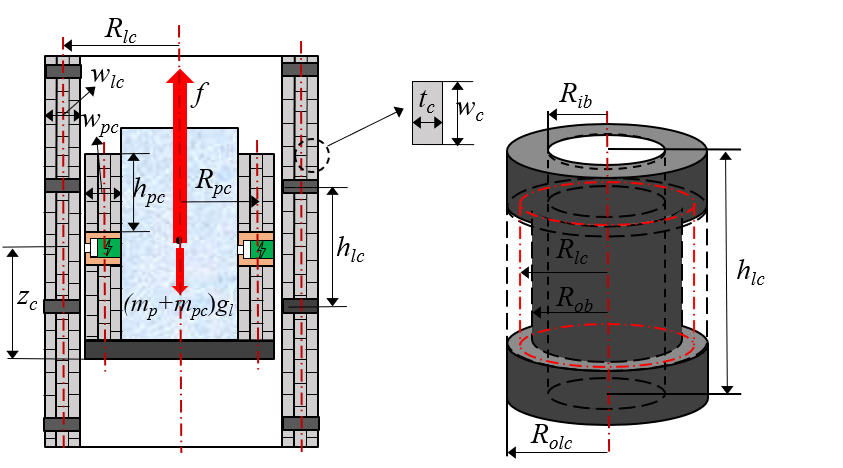}
\caption{The coils winding, the tape, and the bobbin dimensions.}
\label{fig:coils dimensions} 
\end{figure}
\subsection{The required mass and energy of the launch barrel}
In order to generate a large electromagnetic propulsion force, the superconducting coils can be utilized because the electromagnetic force is proportional to the current on the coils winding, while the superconducting coils allow a larger current with less loss. Besides, the lunar temperature can be as low as the operating temperature of the high-temperature superconducting material. The application of the superconducting coils is the key to generate a massive propulsion force. Finally, the superconducting coils are usually shaped like tapes, so we adopt the tape width $w_c$ and the thickness $t_c$ instead of the string radius $r_c$ as the measurements of the coil dimensions. The following derivations are based on the dimensions illustrated in Figure \ref{fig:coils dimensions}.

Given the payload-carrier mass $m_{pc}$, the mass density $\rho_c$, the average radius of the payload-carrier coils $R_{pc}$, the height of the upper or lower payload-carrier coils $h_{pc}$, we can obtain the width $w_{pc}$ or thickness of the payload-carrier coils:
\begin{align} \label{w_pc}
    w_{pc} & =\frac{m_{pc}}{2\rho_c(2\pi R_{pc})h_{pc}}. 
\end{align}\par 
Given the current density $J_c$ and the area $w_c t_c$ of each coil, the current $I_{pc}$ or $I_{lc}$ on each coil can be obtained as $J_c w_c t_c$.
\par Given the height $h_{lc}$, the width $w_{lc}$ of the launch barrel coils winding and the dimensions of each coil $w_c$ and $t_c$, the number of turns $N_{lc}$ on the launch barrel coils can be obtained. The similar calculation can be applied to obtain the number of turns of the coils winding $N_{pc}$ on the upper payload-carrier:
\begin{align}  \label{N_pc,N_lc}
    N_{lc} & = \frac{h_{lc}}{w_c}\frac{w_{lc}}{t_c},N_{pc} = \frac{h_{pc}}{w_c}\frac{w_{pc}}{t_c}.
\end{align}\par 
The width of the launch barrel coils winding $w_{lc}$ needs to be obtained through the formulation of the electromagnetic force $G_f(w_{lc})$. To obtain the width, the function $g(w_{lc})$ needs to be solved. The detailed calculation of the electromagnetic force is derived and presented in the Appendix. $g(w_{lc})$ can be explained that a bigger propulsion force $f$ needs a thicker coils winding $w_{lc}$:
\begin{align} \label{w_pc}
    g(w_{lc}) & = f-2h_{lc}w_{lc}h_{pc}w_{pc}{J_c}^2G_f(w_{lc})=0.
\end{align}\par 

Given the average radius, the width, and the height of the coils winding on the upper payload-carrier and on a sectional launch barrel, the self-inductance of the sectional launcher barrel coils $L_{lc}$ and the upper payload carrier coils $L_{pc}$ can be calculated as \cite{Slobodan2000inductance}: 
\begin{align}
\label{L_lc}
L_{lc}=\frac{\mu_0 \pi N_{lc}^2 R_{lc}}{2\beta_{lc}}\text{T}(k_{Ilc}),~ \\ L_{pc}=\frac{\mu_0 \pi N_{pc}^2 R_{pc}}{\beta_{pc}}\text{T}(k_{Ipc}),
\end{align}
where:
\begin{align} \nonumber
{T}(k_{Ilc}) & = \frac{4}{3\pi\beta_{lc} k_{Ilc}^3}[(2k_{Ilc}^2-1)\text{E}(k_{Ilc}) \\  & +(1-k_{Ilc}^2)\text{K}(k_{Ilc})-k_{Ilc}^3],
~\\  \nonumber
{T}(k_{Ipc}) & = \frac{4}{3\pi\beta_{pc} k_{Ipc}^3}[(2k_{Ipc}^2-1)\text{E}(k_{Ipc}) \\  & +(1-k_{Ipc}^2)\text{K}(k_{Ipc})-k_{Ipc}^3],
~\\ 
k_{Ilc}^2 & = \frac{1}{1+\beta_{lc}^2},~\beta_{lc} =\frac{h_{lc}/2}{R_{lc}}, \\ 
~k_{Ipc}^2  & = \frac{1}{1+\beta_{pc}^2},~ 
\beta_{pc}  =\frac{h_{pc}/2}{R_{pc}},
\end{align}
where $\text{K}$ and $\text{E}$ are the complete elliptic integrals of the first and second kind. $R_{lc}$ and $R_{pc}$ are the average radius of the launch-barrel coils and the payload carrier coils, respectively. $h_{lc}$ and $h_{pc}$ are the height of the sectional launch-barrel coils and the upper (or the lower) payload carrier coils, respectively. $N_{lc}$ and $N_{pc}$ are the number of turns of the coils winding on the sectional launch-barrel and on the upper (or the lower) payload carrier. $\beta$ is an aspect ratio as defined. $\mu_0$ is the space permeability 1.257 $\times$ 10$^{-6}$ H/m. Note that the Eq. (\ref{L_lc}) is utilized providing $R_{olc}/R_{ilc}\approx$1, where $R_{ilc} = R_{lc}-\frac{w_{lc}}{2},~R_{olc} = R_{lc}+\frac{w_{lc}}{2}$, $w_{lc}$ is the width of the launch-barrel coils, $R_{olc}$ and $R_{ilc}$ are the outer and inner radius of the launch-barrel coils. $L_{pc}$ is the total inductance of the upper and lower coils winding connected in series.
\par
The energy of the launcher barrel coils $E_{lc}$ and the energy of the payload carrier coils  $E_{pc}$ can be calculated as:
\begin{align}
\label{E_lc}
E_{lc} & = \frac{1}{2}L_{lc}I_{lc}^2\frac{H}{h_{lc}},~E_{pc}= \frac{1}{2}L_{pc}I_{pc}^2, \\  E_t & =E_{lc}+E_{pc}+E_{ke},
\end{align} 
where $I_{lc}$ or $I_{pc}$ is the current on each coil, $L_{lc}$ and $L_{pc}$ are the self-inductance of the coils winding on sectional launch barrel and on the payload-carrier, $H$ and $h_{lc}$ are the total and the sectional launch barrel height. Note that $E_{lc}$ is only the launch barrel coils energy consumption to generate the required propulsion force. Besides $E_{lc}$ and $E_{pc}$, the total required energy of the launcher $E_t$ also includes the converted kinetic energy $E_{ke}$ of the payload as demonstrated in section 5.4. 
Finally, the mass of the launch barrel coils for this design can be calculated as:
\begin{equation}
     m_{lc} = \rho_c(2\pi R_{lc}) w_{lc} H.
\label{m_lc}
\end{equation}\par
Now, the mass of the coils winding on the launch barrel is provided by Eq. (\ref{m_lc}) based on the required static electromagnetic propulsion force. Since the launch barrel is a high tower subjecting to a large propulsion force (worst at the tip), its supporting structure must be heavy in order to satisfy the Euler buckling formula. It is also important to evaluate the mass of the supporting structure, which will be introduced in Section 4.

\subsection{The LC circuit error}
In this section, we demonstrate how to calculate the required number of ultra-capacitors in the payload-carrier LC circuit and what is the duration of the required current. The same calculations can be applied to obtain the design parameters of the launch-barrel LC circuit.
\par The coils winding on the payload carrier can be powered by ultra-capacitors (super-capacitors) because a large amount of energy needs to be released to the coils in a very quick manner. The ultra-capacitors fit this function very well. Given the required energy of the payload carrier coils $E_{pc}$ and the energy storage capability of each ultra-capacitor $E_{uc}$, the required number of the ultra-capacitor in parallel $N_{sb}$ can be calculated. And the total capacitance $C_{sb}$ and total mass $m_{sb}$ of the super-capacitor bank can be computed as well:
\begin{align}
\label{N_sb}
N_{sb} & = \frac{E_{pc}}{E_{uc}},
~
C_{sb} = \frac{C_{uc}}{N_{sb}},
~
m_{sb} = N_{sb}m_{uc},
\end{align}
where $C_{uc}$ and $m_{uc}$ are the capacitance and the mass of each ultra-capacitor, respectively.
The coils winding on the payload-carrier is an inductance, while the super-capacitor bank is a capacitance. We need to evaluate the LC circuit in series and exam its solution when providing a required current to the coils winding via the super-capacitor bank.  Given the inductance $L_{pc}$ and the capacitance $C_{sb}$, the governing equation of the current $I(t)$ (with initial conditions) is:
\begin{align}
\label{I(t)}
& \frac{d^2I(t)}{dt^2}+ \frac{1}{L_{pc}C_{sb}}I(t) = 0,~I(0)=0,~ \\ 
& V(0) =L_{pc}I_0\omega_0=N_{sb}V_{uc}.
\end{align} 
The solution of this differential equation is:
\begin{align}
\label{I0}
I(t) = I_0\text{sin}(\omega_{0}t),~\omega_0  =\frac{1}{\sqrt{L_{pc}C_{sb}}}, ~ I_0 = \frac{N_{sb}V_{uc}}{\omega_0L_{pc}},
\end{align}
where $\omega_0$ and $I_0$ are the current natural frequency and magnitude of the LC circuit, respectively.
\par In order to provide the designed current $I_{pc}$ to the payload carrier coils, we need to make sure the super-capacitor bank can provide the required current within the launch time, that is  the duration of the provided current $\Delta T$ is much larger than the launch time $T$. Assuming that at time $t_1$ and $t_2$, the current $I$ is $I_{pc}$, we can solve for $t_1$ and $t_2$, and then calculate the designed current duration, which is given by:
\begin{align}
t_1 & = \frac{\text{sin}^{-1}(\frac{I_{pc}}{I_0})}{\omega_{0}},~t_2=\frac{\pi-\text{sin}^{-1}(\frac{I_{pc}}{I_0})}{\omega_{0}}.
\end{align}
Duration of the required current is $\Delta T= t_2-t_1$, the error of this approximate current is $\epsilon=1-\frac{I_0}{I_{pc}}$. 
\subsubsection{The LC circuit on the launch barrel}

\section{Tensegrity tower for launching} 

\begin{figure}
  \centering
\begin{subfigure}{0.45\textwidth}
    \centering
    \includegraphics[scale=0.25]{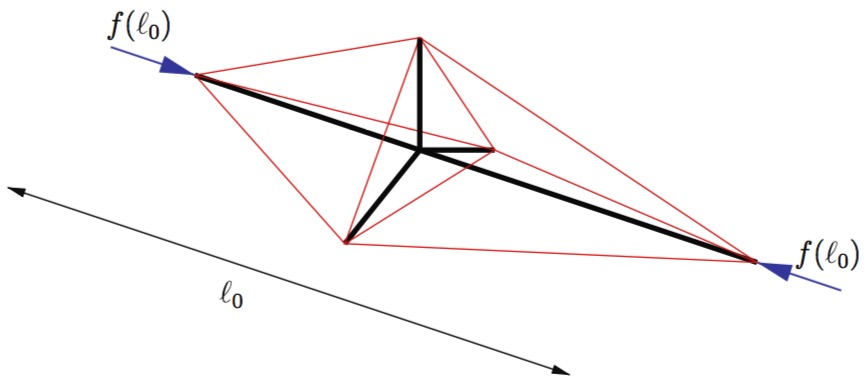}
    \caption{3D T-Bar unit.}
     \label{3DTbar}
\end{subfigure}
\begin{subfigure}{0.45\textwidth}
    \centering
    \includegraphics[scale=0.45]{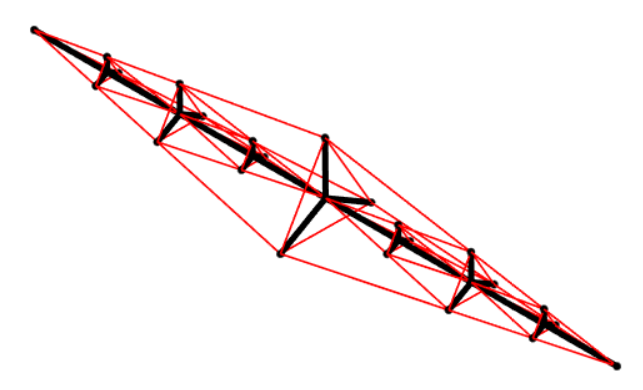}
    \caption{Three-dimensional $T_3$ structure.}
     \label{3d_tbar_system}
\end{subfigure}
\caption{Three-dimensional tensegrity T-Bar unit and T$_3$-Bar structure. Black and red lines are bars and strings. The T-Bar angle $\alpha$ is the angle between the outer strings of the T-Bar unit and its horizontal line.}
\end{figure}

\begin{figure}
\centering
\includegraphics[scale=0.65]{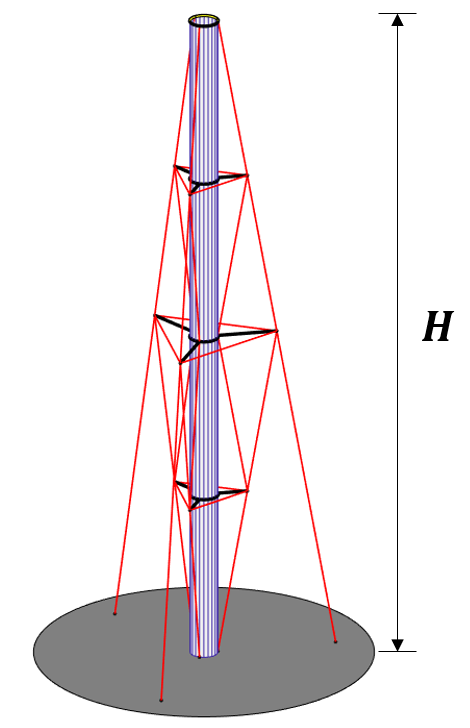}
\caption{The electromagnetic tensegrity lunar launcher, 3D T-Bar tensegrity structure where black lines are bars, red lines are strings, and the light blue cylinder is the launch barrel.}
\label{launcher_round2} 
\end{figure}

For launching operations, the launching reaction force is a big compressive load on the supporting structure. To save cost, it is critical to design a mass-efficient structure. It has been proved by Skelton and de Oliveira that T-Bar structures require less mass than a continuum bar in taking the same compression load (propulsion force in our case) based on the Euler buckling formula. A three-dimensional T-Bar unit with length $l_0$ is shown in Figure~\ref{3DTbar} \cite{skelton2009tensegrity}. Each longitudinal bar in the T-bar structure can be replaced by another T-bar tensegrity unit while preserving the total length of the structure, which is called a \textit{self-similar rule}. The iteration of the self-similar process times $i$ is defined as the \textit{complexity of the structure}. Figure~\ref{3d_tbar_system} shows a T-Bar structure of complexity $i=3$. To accommodate the launching application, we take the half part of the T-Bar structure in Figure~\ref{3d_tbar_system} and replace the continuous middle bar with a hollow pipe (the launch barrel in our case). The final configuration of the launcher is shown in Figure~\ref{launcher_round2}.

%Let the structure complexity be $i$, we can see that the vertical launch barrel is composed of 2$^{i}$ numbers of sectional launch barrels with length $l_i$, horizontal bars $l_{vi}$, and strings $s_i$. The payload mass can be viewed as a compressive load $f$ on the top of the tower. The height of the tower is $H$. For the minimum mass tensegrity design, we need to solve for the optimal self-similar iteration value $i^{*}$ by Section \ref{design_algorithm}. A detailed example is shown in Section \ref{tw_design}.}

\subsection{Tensegrity structure notations}

We define $N = \begin{bmatrix} \bm{n}_1 & \bm{n}_2 & \cdots & \bm{n}_{n} \end{bmatrix} \in \mathbb{R}^{3 \times n}$ is the nodal matrix with each column of $N$ to represent the node position of each node, $n$ is the number of nodes. $C_s\in \mathbb{R}^{\alpha \times n}$ and $C_b \in \mathbb{R}^{\beta \times n}$ are the connectivity matrices of strings and bars (with 0, -1, and 1 contained in each row), respectively. The number of bars and strings are denoted by $\alpha$ and $\beta$. The external force matrix $W=\begin{bmatrix} \bm{w}_1  & \bm{w}_2 & \cdots & \bm{w}_{n} \end{bmatrix} \in \mathbb{R}^{3\times n}$ contains each of its column as the force vector acting on the corresponding node, $\gamma \in \mathbb{R}^{\alpha}$ is a vector of force densities (force per unit length) in the strings, and $\hat{\vec{v}}$ is a diagonal matrix of the elements of a vector \vec{v}. The string and bar vectors are contained in the string matrix $S= \begin{bmatrix} \bm{s}_1  & \bm{s}_2 & \cdots & \bm{s}_{\alpha} \end{bmatrix}\in \mathbb{R}^{3\times \alpha}$ and in the bar matrix $B= \begin{bmatrix} \bm{b}_1 & \bm{b}_2 & \cdots & \bm{b}_{\beta} \end{bmatrix} \in \mathbb{R}^{3\times \beta}$ respectively. From the definitions we also have $S = NC_s^T$ and $B = NC_b^T$.

% \subsection{Tensegrity structure design}

% {\color{blue}For description convenience, we also make labels for the structure members, as shown in Figure \ref{3D_T_Bar_TS}. Usually, the tensegrity structure is composed of vertical bars, horizontal bars, and strings. But the vertical bars in our case are replaced with sectional launch barrels, the mass of which is evaluated in Section 3. Therefore, the tensegrity supporting structure, in this case, is composed of 2$^{i}$ numbers of sectional launch barrels with length $l_i$ (the inner radius needs to be evaluated), solid horizontal bars $l_{vi}$ (thick lines), and strings $s_i$ (thin lines). The launch barrel outer radius is dictated by the outer radius of launch-barrel coils $R_{olc}$ as shown on the RHS in Figure \ref{fig:coils dimensions}. The horizontal bars $l_{vi}$ are subject to compressive forces $f(l_{vi})$ to prevent sectional launch barrels from global buckling under force $f$. The strings $s_i$ are subject to tension $t(s_i)$ to balance the force of the whole tensegrity structure. The height of the tensegrity supporting structure is dictated by the launch barrel height $H$. 

%In order to calculate the minimum mass of the tensegrity structure to support the launching barrel coils, we need to solve for the optimal self-similar iteration $i^{*}$, where $i$ is called the number of self-similar iterations such that the launch barrel will be cut into 2$^{i}$ pieces. } {\color{red}Note: what is the relation of n and i, $\alpha$, $\beta$.where is t(si)..., what is the class of the tensegrity tower,}

\subsection{Minimal mass tensegrity design}

In this section, we present an algorithm in designing the structure member thickness to achieve structure minimal mass subject to the load requirements based on the proposed topology. The idea is that the static equilibrium equation can be written in terms of force density $x$ of the structure members. Minimal mass design is to calculate the critical mass that every structure member fails at the same time, which is also a function of force density $x$ (bar mass subject to buckling is nonlinear in its force density). Structure gravity can be viewed as an external force on the structure, but it is also coupled with structure mass. Thus, the minimal mass tensegrity design problem yields to find the optimal force density $x$ for given loads and topology of the structure, which can be formulated as a nonlinear optimization problem.

% We will show that the solution yields a 

% yields a nonlinear optimization problem.

\subsubsection{Static equilibrium}

The static equilibrium equation for a given tensegrity structure and given external force can be written as \cite{chen2020general}:
\begin{align}\label{equality}
Ax & = W_{vec}, ~x = \begin{bmatrix} \gamma^T & \lambda^T \end{bmatrix}^T,
\end{align}
where:
\begin{align}\label{A_w}
A(i,:) = S\reallywidehat{(C_s \bm{e}_i)} -B \reallywidehat{( C_b \bm{e}_i)}, ~W_{vec}(i,:) =W \bm{e}_i,
\end{align}
where $\bm{e}_i = \begin{bmatrix}0 & 0 & \cdots & 1 & \cdots & 0 & 0 \end{bmatrix}^T$ is a column with 1 as the $i${th} element and zeros elsewhere, $\gamma$ and $\lambda$ are force densities in the strings and bars.

\subsubsection{Mass of structures}

Since the hollow bars require less mass than a solid bar when subject to buckling. The mass formula for the hollow bar tensegrity system is given by \cite{chen2020general}:
\begin{align}
\nonumber
    M =  & \frac{\rho_s}{\sigma_s}(vec{(\lfloor S^TS \rfloor))^T}\gamma + \frac{\rho_b}{\sigma_b}(vec{(\lfloor B^TB \rfloor(I-Q)))^T}\lambda \\ \nonumber
    & + \frac{\rho_b (vec{(\lfloor B^TB \rfloor^\frac{1}{2}Q))^T}}{\sqrt{\pi E_b}} \\ \label{hollow_yb_case} &  (\sqrt{\pi^3 E_b r_{in}^4 + 4(\lfloor B^TB \rfloor^\frac{3}{2}Q)\lambda}  -\pi r_{in}^2 \sqrt{\pi E_b}),
\end{align}
where $r_{in} = \begin{bmatrix} r_{1_{in}} & \cdots & r_{j_{in}} & \cdots & r_{\beta_{in}}\end{bmatrix}^T$ is a vector of the inner radius of all the hollow bars. If one need to use a solid bar in the $j$th bar, we just need to set $r_{j_{in}} =0$. $\rho_s$, $\rho_b$, $\sigma_s$, $\sigma_b$ are the density and yield strength of strings and bars, respectively. The length of each string and each bar are denoted by $||s_i||$, and $||b_j||$ for $i = 1, 2, \cdots, \alpha$; and  $j = 1, 2, \cdots, \beta$, and $E_b$ is Young's modulus of bars. $\lfloor \bullet \rfloor$ is an operator taking the diagonal elements of a matrix, $vec(\bullet)$ is an operator taking the elements of the matrix and form a vector,  $Q\in \mathbb{R}^{\beta \times \beta}$ is a square matrix with 0s or 1s in its diagonal and 0s elsewhere, where the diagonal indices of the 1s denotes the buckling bars, and the diagonal indices of the 1s of $(I-Q) \in \mathbb{R}^{\beta \times \beta}$ labels the yielding bars.

\subsubsection{Gravity of the structure}

Since gravity is unlike a given set of specific external forces that apply to the structure, it is determined by the mass of the structure itself. In other words, the statics mass optimization process is coupled with the gravity force. The total force $W$ can be separated into two parts $W = W_e + W_g$, where $W_g$ is the gravity force, and $W_e$ is other applied external force. The acceleration due to gravity is defined as $\bm{g}$ with lunar gravity as $\bm{g}_{moon} = \left[ \begin{matrix} 0& 0& -1.62 \end{matrix}\right]^T$ m/$s^2$. The gravity force can be modeled by lumped forces equally distributed on the member nodes \cite{chen2020general}. Thus, the gravitational force due to bars and strings can be expressed as:
\begin{align}
\nonumber
  W_{gi}  & =  \frac{1}{2}\bm{g} \frac{\rho_s}{\sigma_s}(vec{(\lfloor S^TS \rfloor))^T}\reallywidehat{|C_s e_i|}\gamma \\ \nonumber  & +  \frac{1}{2}\bm{g}\frac{\rho_b}{\sigma_b}(vec{(\lfloor B^TB \rfloor(I-Q)))^T}\reallywidehat{|C_b e_i|}\lambda \\  \nonumber
  &+ \frac{1}{2}\bm{g}\frac{\rho_b (vec{(\lfloor B^TB \rfloor^\frac{1}{2}Q))^T}}{\sqrt{\pi E_b}}\reallywidehat{|C_b e_i|} \\ & (\sqrt{\pi^3 E_b r_{in}^4 + 4(\lfloor B^TB \rfloor^\frac{3}{2}Q)\lambda} -\pi r_{in}^2 \sqrt{\pi E_b}),
\end{align}
where $|\bullet|$ is an operator getting the absolute value of each element for a given matrix. Stacking all the columns, we get:
 \begin{align}\label{gravity_vec}
    W_{g vec} =  \begin{bmatrix}  W_{g1}^T &  \cdots &  W_{gi}^T & \cdots &  W_{gn}^T \end{bmatrix}^T.
\end{align}

\subsubsection{Stiffness of the structure}

\noindent The  stiffness matrix $K_n$ for tensegrity structure subject to yielding and buckling constraints is given by: $K_n vec(dN) = vec(dW)$, where:
\begin{align} \nonumber
K_n & = (C_s^T \otimes I_3)\textbf{b.d.($K_{s1}$,$\cdots$,$K_{s\alpha}$})(C_s \otimes I_3) \\ \label{Kn_matrix} &
-  (C_b^T \otimes I_3)\textbf{b.d.($K_{b1}$, $\cdots$, $K_{b\beta}$})(C_b \otimes I_3),
\end{align}
and
\begin{align}
    K_{si} & = \gamma_i (I_3 + \frac{E_{si}}{\sigma_s}\frac{s_i s_i^T}{||s_i||^2}) ,\\  \nonumber
       K_{bj} & = \lambda_j (I_3 - (1-Q_{jj})\frac{E_{bj}}{\sigma_b}\frac{b_j b_j^T}{||b_j||^2})  - Q_{jj} \frac{E_{bj}}{\sqrt{\pi E_{bj}}}\\ & (\sqrt{\pi^3 E_{bj} r_{j_{in}}^4 + 4\lambda_j||b_j||^3} -\pi r_{j_{in}}^2 \sqrt{\pi E_{bj}})\frac{b_j b_j^T}{||b_j||^3},
\end{align}
where $\otimes$ is the Kronecker product operator. 

\subsubsection{The structure design algorithm}
\label{design_algorithm}

\noindent The minimal mass problem is formulated as:
\begin{equation}\label{solid_yb_min}
\left \{
\begin{aligned}
& \underset{x}{\text{minimize}}
& & M \\
& \text{subject to}
& & Ax = W_{e~vec} + W_{g~vec}, ~ x \geq \epsilon_0, \text{and}~ \\ & & & eig(K_n) > \mu I
\end{aligned}
\right.,
\end{equation}
where $\epsilon_0$ is the prestress assigned to the strings, and $\epsilon_0 \geq 0$ guarantees that all strings are in tension and all bars in compression, $eig(K_n)$ returns the eigenvalues of the matrix $K_n$, and the system is globally stable at the equilibrium for $\mu \geq 0$. One cannot exactly tell $Q$ for any structure in advance because it is determined by structure topology and external force. To obtain a global solution, the nonlinear optimization problem can be solved in an iterative manner as described in Algorithm~\ref{alg1}.

\begin{algorithm}[h!]
\caption{Minimal mass tensegrity subject to stability and gravity}
{\bf 1)} Given tensegrity structure topology $N$, $C_b$, $C_s$ and external force $W$, compute $A$ and $W_{e~vec}$ from Eq. (\ref{A_w}).\\
{\bf 2)} Let $Q={I}^{\beta \times \beta}$, $W_{g~vec} = 0$, $\epsilon_0 = 0$, $\delta \epsilon = 0 $, $\mu = 0.01$.\\
{\bf 3)} Compute force densities $x$: \\
\While{$\min\{eig(K_n)\}<\mu$}{
\While{$Q_{i+1} \neq Q_{i}$}{
\begin{align}\nonumber 
&\begin{cases}
\underset{x}{\text{minimize}}~
M \\ \text{subject to}~
Ax = W_{e~vec} + W_{g~vec}, ~ \\
~~~~~~~~~~~~~~~~~~~ x \geq \epsilon_0. 
\end{cases} \\ \nonumber 
& \text{Compute $\lambda$ from $x$, update $Q$.}\\ \nonumber
& \text{Update $W_{g~vec}$ from Eq.~(\ref{gravity_vec})}.
\\ \nonumber
& i \leftarrow i+1.
\end{align}
}
Compute stiffness matrix $K_n$ from Eq.~(\ref{Kn_matrix}).\\
$\epsilon_0 \leftarrow \epsilon_0 + \delta \epsilon$.}.
\label{alg1}
\end{algorithm}

\begin{figure}
\centering
\includegraphics[scale=0.4]{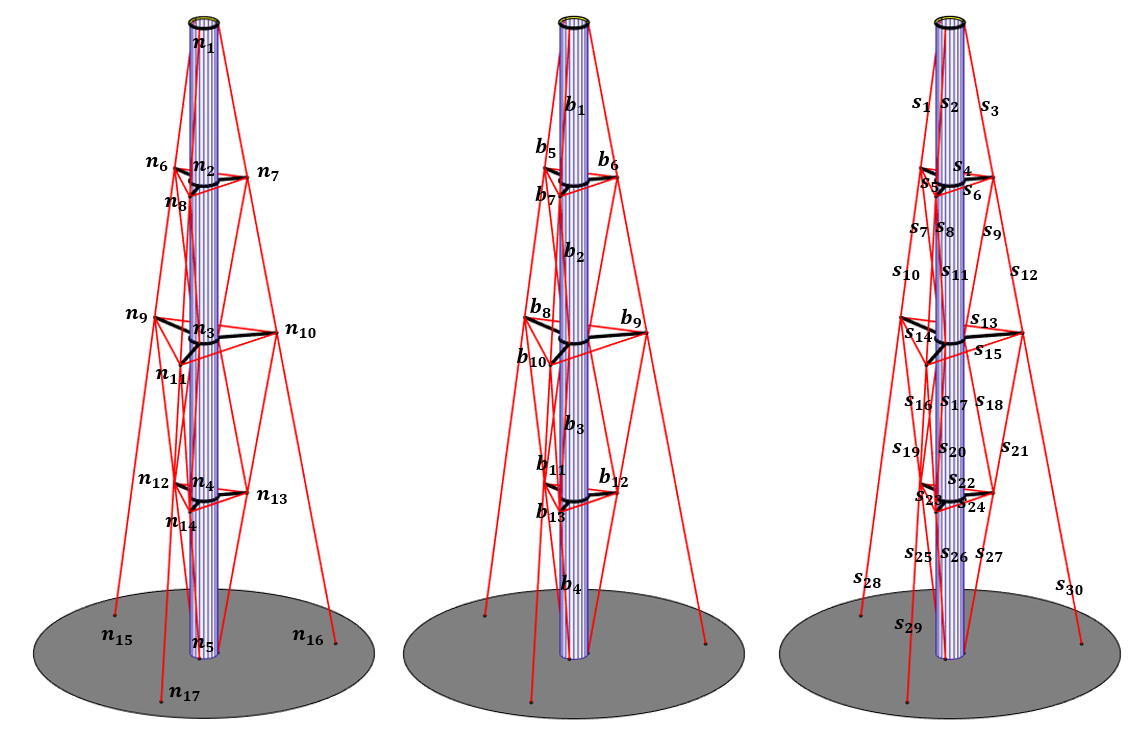}
\caption{Node, bar, and string notations of the tensegrity tower.}
\label{notation_tower} 
\end{figure}

% \begin{figure}
%     \centering
%     \begin{subfigure}{0.45\textwidth}
%     \includegraphics[scale=0.5]{launcher.png}
%     \caption{3D T-Bar tensegrity structure.}
%     \label{fig:tower}
%     \end{subfigure}
% \begin{subfigure}{0.45\textwidth}
%     \centering
%     \includegraphics[scale=0.57]{3D_T_Bar_TS.png}
%     \caption{3D T-Bar tensegrity structure notations.}
%      \label{3D_T_Bar_TS}
% \end{subfigure}
% \caption{The electromagnetic tensegrity lunar launcher, 3D T-Bar tensegrity structure where black lines are bars, red lines are strings, and the light blue cylinder is the launch barrel. {\color{red}Label all the nodes, bars, and strings in (b), and how the bar is generated by the nodes at its two ends. Put height $H$ and $f$ in the algorithm.}}
% \end{figure}

\section{Case Study}
\subsection{Prerequisite}
\subsubsection{Electromagnetic Propulsion}
The launch prerequisite are introduced here:
1. The launch kinematics, which includes the lunar exit velocity $v_e$, the acceleration $a_1$ (determined by payload material strength), payload mass $m_p$, payload-carrier mass $m_{pc}$.
2. The coils winding installation requirements, which are the average radius of the launch barrel and the payload carrier, $R_{lc}$ and $R_{pc}$; the height of the sectional launch barrel and the upper carrier coils, $h_{lc}$ and $h_{pc}$; the center distance of the sectional launch barrel and the corresponding upper carrier coils $z_c$.
1 and 2 are listed in Table \ref{tab:1}.
\par
For a preliminary mass and function design, the specifications of some products applied in this study are also introduced here: 1. Superconducting tape wire whose type is SuperPower$^@$ 2G high temperature superconducting (HTS) tape wire SCS4050 \cite{SuperPower}. The width $z_c$, the thickness $w_c$, the current density $J_c$ and the mass density $\rho_c$.
%  (high current density achieves in 70 K)
2. Ultra-capacitors, which are the Maxwell Technologies$^@$ 2.7 V 350 F ultra-capacitor cells BCAP0350 \cite{Supercapacitor}. the capacitance, the Equivalent Series Resistance (ESR), the fully charged voltage, the energy storage and the mass of each ultra-capacitor, $C_{uc}$  $R_{uc}$, $V_{uc}$, $E_{uc}$ and $m_{uc}$. These are listed in Table \ref{tab:2}. Note that the operation condition of the ultra-capacitors is at room temperature, so the isothermal coatings should be installed for lunar applications.

\begin{table}[width=.9\linewidth,cols=2,pos=ht]
\caption{Launch kinematics and the installation requirements}
\begin{tabular*}{\tblwidth}{@{} LLLL@{} }
\toprule
\multicolumn{2}{c}{}\\
$v_e$ [m/s] & 2,200 & $a_1$ [m/s$^2$] & 30,000\\ \midrule
$m_p$ [kg] & 1,500 &  $m_{pc}$ [kg] & 50\\
\midrule
$z_c$ [m] & 0.5 & $h_{pc}$ [m] & 0.5\\
\midrule
$h_{lc}$ [m] & 1.0 &  $R_{pc}$ [m] & 0.5\\
\midrule
$R_{lc}$ [m]  & 0.6 & $y_c$ [mm] & 5.0\\
\bottomrule
\end{tabular*}
\label{tab:1}
\end{table}

\begin{table}[width=.9\linewidth,cols=2,pos=ht]
\caption{Specifications of the applied products}
\centering
\begin{tabular*}{\tblwidth}{@{} LLLL@{} }
\toprule
\multicolumn{2}{c}{}\\
$J_c$, [A/mm$^2$] & 325 (70 K) & $\rho_{c}$, [kg/m$^3$]& 6,300\\ \midrule
$w_c$, [mm]& 4.0 & $t_c$, [mm]& 0.1\\ \midrule $m_{uc}$, [g] & 65 &$C_{uc}$, [F]& 350\\ \midrule
$V_{uc}$, [V] &2.7& $R_{uc}$, [m$\Omega$] &3.2\\ \midrule
$E_{uc}$, [Wh] & 0.35\\
\bottomrule
\end{tabular*}
\label{tab:2}
\end{table}
\begin{table}[width=.9\linewidth,cols=2,pos=ht]
\caption{Material properties of the structure members (hollow\slash solid bars and strings).}
\centering
\begin{tabular*}{\tblwidth}{@{} LLLL@{} }
\toprule
\multicolumn{2}{c}{}\\
$\rho_h$, [kg/m$^3$] &8,000 & $E_h$,  [GPa] &190\\\midrule
$\sigma_h$, [MPa] & 205 &$\rho_b$, [kg/m$^3$] & 1,500\\\midrule
$E_b$, [GPa] &138 & $\sigma_b$, [GPa] & 1.72\\\midrule
$\rho_s$, [kg/m$^3$] &970 &
$E_s$, [GPa] & 120 \\\midrule
$\sigma_s$, [GPa] & 2.7\\\bottomrule
\end{tabular*}
\label{tab:5} 
\end{table}
% {\color{red}
\subsubsection{Tensegrity Tower Supporting Structure}
% {\color{blue}
% Things prepare for calculation. 

% Table 3. gives the relationship of $W$ and $f$ , i.e., $W$(3,\text{top node index}) = $-f$ (3 means z-coordinate), $N$, $C_b$, $C_s$ of the case structure. One node and bar/string to show how we formulate the matrix. Figure to show.  Some of the parameters below can be put into the Table.
% }

Based on the propulsion forces obtained by the previous sections, now we can design the structures: complexity and thickness of all the structure members. The horizontal bars and the strings are made of carbon rods and ultra-high-molecular-weight polyethylene (UHMWPE). To prevent the whole structure from global buckling, we choose the T-Bar structure angle and a prestress on each string to be $\alpha_i=\alpha= \frac{\pi}{18}$ and $t(s_i) = t=0.01f$ as a demonstration. Assuming propulsion force $f$ = 46.5 MN, the launch-barrel height $H$ = 83.24 m, and the outer radius of the launch barrel $R_{olc}$ = 620.3 mm. Tensegrity structure material properties we choose are listed in Table \ref{tab:5}, $\rho$ represents the mass density, $\sigma$ is the yield strength, $E$ is Young's modulus, subscripts $h$, $s$, and $b$ represent hollow bars, strings, and horizontal solid bars, respectively. The coils winding outer radius of the launch barrel $R_{olc}$ determines the hollow bobbin's outer radius. The inner radius of the bobbin is obtained via Eq. (26) based on the objective function Eq. (20).
% }

The stainless steel bobbin is commonly used \cite{wang2013practical}. As the austenitic stainless steel and the air has a very similar permeability with the free space permeability, so we assume that the stainless steel bobbin would not change the magnetic field between the two coils resulting in the same mutual electromagnetic static force formula developed between two air-cored coils. Note that the design in this paper is based on the static electromagnetic force. If dynamic electromagnetic losses are to be considered in terms of the skin effect, lizt wire technology might be applied to mitigate the skin effect for frequencies of a few kilohertz to about one megahertz.

\subsection{Design results}
\begin{table}[width=.9\linewidth,cols=2,pos=ht]
\caption{Kinematics and geometry design results}
\centering
\begin{tabular*}{\tblwidth}{@{} LLLL@{} }
\toprule
\multicolumn{2}{c}{}\\
$f$, [MN]& 46.5 & $a_2$, [m/s$^2$]& 9.34 $\times$ 10$^5$\\ \midrule
$H$, [m]& 83.24 & $w_{pc}$, [mm] & 2.5 \\ \midrule
$N_{lc}$ & 1.02 $\times$ 10$^5$ &$N_{pc}$ & 3,125 \\ \midrule
$w_{lc}$, [mm] & 40.6 &$f_s$, [N] & 2.09 $\times$ 10$^3$\\ \midrule
 $R_{ilc}/R_{olc}$, [mm]&579.7/620.3& $m_{lc}$, [kg] & 8.03 $\times$ 10$^4$\\ \bottomrule
\end{tabular*}
\label{tab:3} 
\end{table}
\subsubsection{Electromagnetic Propulsion}

Given the launch kinematics, the propulsion force $f$, the deceleration $a_2$, the launch barrel height $H$, and the total launch time $T$ can be obtained via Eqs. (1)-(4). The self-inductance, the energy of the carrier coils and the sectional launch barrel coils, $L_{pc}$, $L_{lc}$, $E_{pc}$ and $E_{lc}$ can be obtained via Eqs. (5)-(12). The total energy $E_t$ can be calculated via Eq. (12). The capacitance $C_{sb}$, the mass $m_{sb}$ and the required number $N_{sb}$ of the ultra-capacitor bank can be obtained via Eq. (14). The discharge time of the ultra-capacitor bank $\Delta T$ and the error of the approximate current $\epsilon$ can be calculated through Eqs. (15)-(17). The results are shown in Table \ref{tab:3} and Table \ref{tab:4}. To summarize, given launch mass (a 1,500 kg payload installed in a 50 kg payload-carrier), exit velocity 2,200 m/s, and launch acceleration 30,000 m/s$^2$, the required launch barrel height is 83.24 m, launch time is 0.08 s, propulsion force is 46.5 MN, and restoring force is 2.09 $\times$ 10$^3$ N, considering a 5 mm misalignment of the two sets of concentric coils. The required energy of the launch-barrel coils and the payload carrier coils are 1,852.2 kWh and 46.9 Wh, respectively. The total energy is 2860.5 kWh. In addition, lunar surface radiation power is 342 W/m$^2$, a 2860.5 kWh energy would take a 250 m$^2$ solar panels with 20\% efficiency (68 W/m$^2$) about 170 hrs to get the power/energy storage system fully charged. Finally, the designed current in a tape wire is 130 A (provided by 134 ultra-capacitors), and the LC circuit has a capacity to supply 130 A during 0.08 s with an error of 0.07\%.\par
% {\color{red}

\subsubsection{Tensegrity Tower Supporting Structure}

% {\color{blue} Update hollow bar inner radius and total weight.}

Based on the algorithm developed in Section \ref{design_algorithm}, the optimal complexity $i^*$ of the tensegrity structure subject to buckling and yielding is 3. The minimum mass tensegrity launcher structure itself is 3.84 $\times$ 10 $^4$ kg, where the mass of the solid bars, strings, and hollow bars are 3.18 $\times$ 10$^3$ kg, 3.82 $\times$ 10$^3$ kg, and 3.14 $\times$ 10$^4$ kg.
% }

% the LC circuit can provide a current as high as 129.91 A. The required current discharging time of the supercapacitor bank is 0.65 s (more than enough to cover the launch time 0.08 s), the error of this approximate  current is 6.9e-4.\par

%$a_2$, m$\cdot$\text{s}$^{-2}$& $H$, m&  $w_{pc}$, mm &$N_{lc}$ &$N_{pc}$ & $w_{lc}$, mm & $f_s$, N  & $m_{lc}$, kg&$R_{ilc}/R_{olc}$, mm\\\hline
%4.65e7 & 9.34e5 & 83.24 & 2.5 & 1.02e5 & 3125 & 40.6 & 2.09e3 & 8.03e4&579.7/620.3\\\hline

\begin{table}[width=.9\linewidth,cols=2,pos=ht]
\caption{Circuit and energy design results}
\centering
\begin{tabular*}{\tblwidth}{@{} LLLL@{} }
\toprule
\multicolumn{2}{c}{}\\
$T$, [s]&0.08 & $L_{lc}$, [H]&9.48 $\times$ 10$^3$ \\\midrule
 $E_{lc}$, [kWh]&1,852.2 & $E_{pc}$, [Wh] & 46.9 \\\midrule
$N_{sb}$ &134 &$C_t$, [F] & 2.61 \\\midrule
$R_t$, [m$\Omega$] & 2.39 $\times$ 10$^{-2}$ & $\Delta T$, [s] & 0.65 \\\midrule
$I_{pc}$/$I_{lc}$, [A] & 130 & $m_{sb}$, [kg] & 8.7\\\midrule
$L_{pc}$, [H] & 20.27& $\epsilon$ & 0.07\% \\\midrule $E_t$, [kWh] & 2860.5\\
\bottomrule
\end{tabular*}
\label{tab:4}
\end{table}

\subsection{Discussion on the mass of the launch-barrel coils}
To get an idea of how the mass of the launch-barrel coils, $m_{lc}$, varies with accelerations, $a_1$, we can compare the mass of the launch-barrel coils calculated from the numerical simulations (solid blue lines) with that from the approximations (red dotted lines), providing the launch kinematics and installation requirements.\par
\begin{figure}
        \centering
        \includegraphics[scale=0.22]{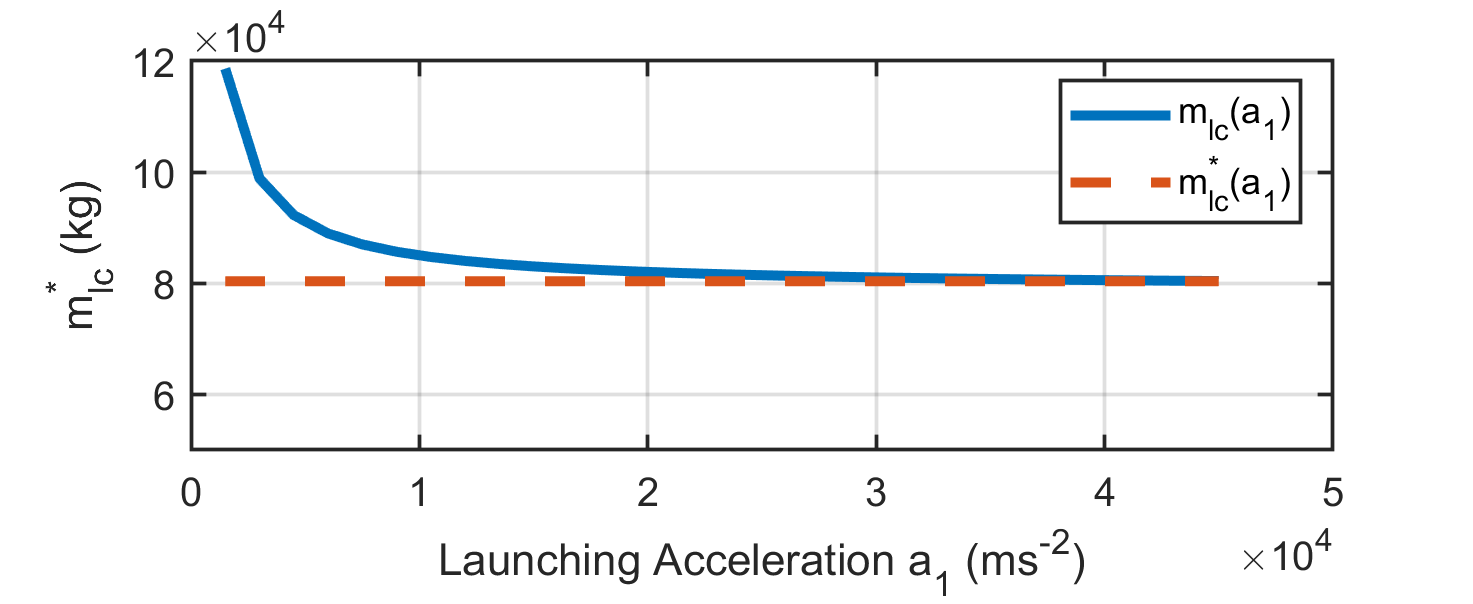}
        \caption{Comparison of $m_{lc}(a_1)$ calculated from the simulation with that from the analytic approximation. }\label{fig:m_lc(a1)}
    \end{figure}
\begin{figure}
\centering
     \begin{subfigure}[b]{0.45\textwidth}
         \centering
        \includegraphics[width=1.1\linewidth]{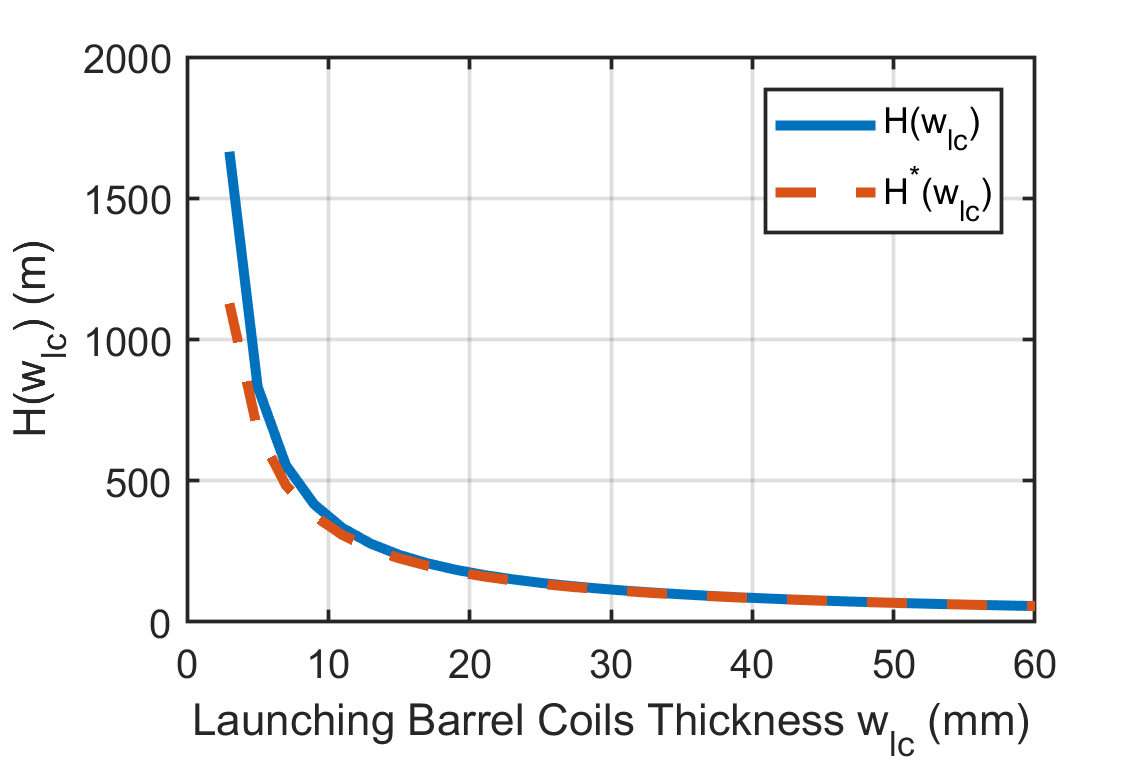}
        \caption{Comparison of $H(w_{lc})$ calculated from the simulation  with that from the analytic approximation. }\label{fig:H(w)}
    \end{subfigure}
     \begin{subfigure}[b]{0.45\textwidth}
         \centering
        \includegraphics[width=1.08\linewidth]{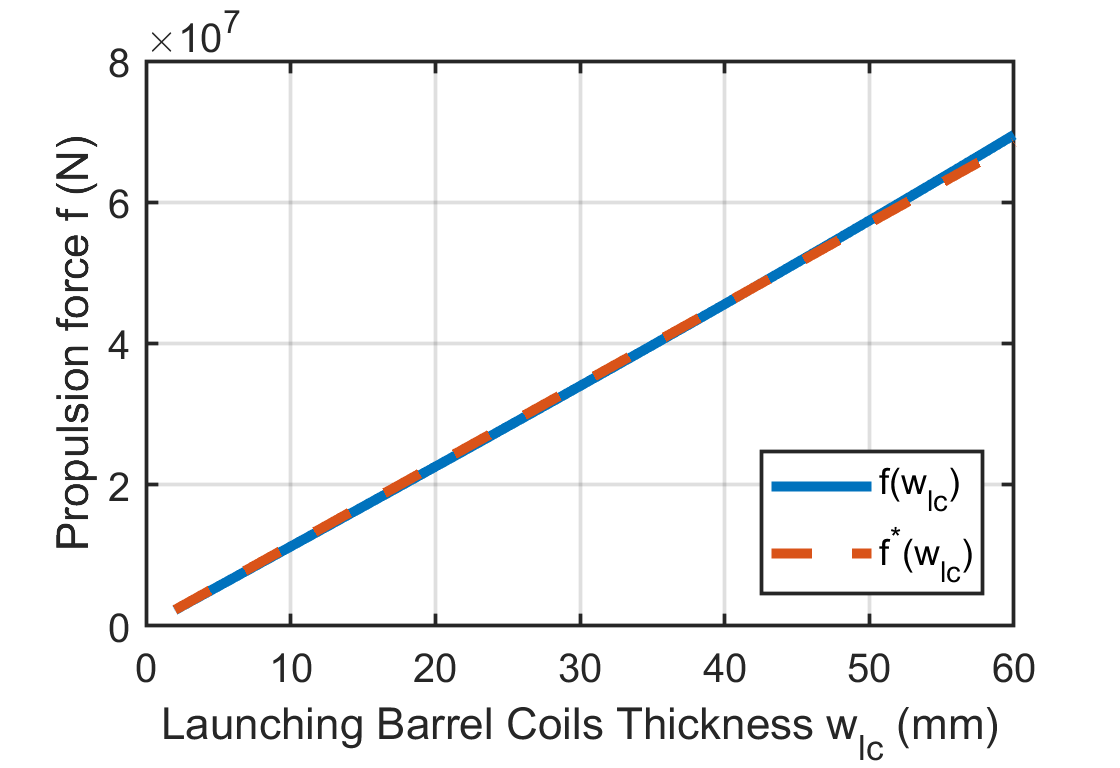}
        \caption{Comparison of $f(w_{lc})$ calculated from the simulation  with that from the analytic approximation.}\label{fig:f(w)}
           \end{subfigure}
           \caption{Comparison of the analytical approximation and the simulation on the launch barrel height and the propulsion force.}
\end{figure}

For analytic approximation, the unit propulsion force $G_f$ is approximately 4.33$\times\text{10}^{-\text{6}}$ N/A$^2$ via Eq. (\ref{eqn:G_f}) based on thin wall assumption ($w_{lc}<0.1R_{lc}$) \cite{babic2008magnetic}. Therefore, the mass of the launch barrel coils $m_{lc}$ is approximately a constant function of launching acceleration $a_1$ as shown in Eq. (\ref{m_lc_new}) and the dotted line in Figure \ref{fig:m_lc(a1)}. The constant value is 8.0 $\times$ 10$^4$, providing Eq. (\ref{H(w_lc)}) and Eq. (\ref{f(w_lc)}). These two equations mean that the launch-barrel height is a reciprocal function of the propulsion force/acceleration as shown in Figure \ref{fig:H(w)} and the propulsion force is a linear function of the launch-barrel coils as shown in Figure \ref{fig:f(w)}.

For numerical simulations,the launch-barrel coils mass $m_{lc}$, asymptotically decreases to 8.0 $\times$ 10$^4$ kg as the launching acceleration $a_1$, increases to 4.5 $\times$ 10$^4$ m/s$^2$ as shown in the solid blue line in Figure \ref{fig:m_lc(a1)}. It is worth noting that these conclusions are based on the assumption that the mutual inductance between the launch-barrel coils unit and the payload-carrier coils unit does not depend on $w_{lc}$ with the thin wall solenoid assumption.
\begin{align}
\nonumber
    m_{lc}^* & \approx \rho_c (2\pi R_{lc})w_{lc}H^*(w_{lc}) = \rho_c (2\pi R_{lc})  \\ \label{m_lc_new} & \frac{v_e^2(m_p+2m_{pc})}{4h_{pc}^* h_{lc}^* w_{pc}^* J_c^2 G_f^*},\\
\label{H(w_lc)}
    H^*(w_{lc}) &  = \frac{v_e^2(m_p+2m_{pc})}{2f^*(w_{lc})}\approx\frac{v_e^2(m_p+2m_{pc})}{4h_{pc}^*h_{lc}^*w_{pc}^*w_{lc}J_c^2G_f^*},\\
\label{f(w_lc)}
    f^*(w_{lc}) & \approx 2h_{pc}^*h_{lc}^*w_{pc}^*w_{lc}J_c^2G_f^*,
\end{align}
where $*$ denotes the case study parameters.

\par 
% because We can know that the mass of the launch-barrel coil is approximately a constant (also shown in Figure \ref{fig:m_lc(a1)} the dotted line, where the actual value is in the solid line)

% caption{Illustration of the approximate functions $f^*(w_{lc})$ and $H^*(w_{lc})$.}\label{fig:3section{Tensegrity Structure}
\subsection{Discussion on energy and efficiency}
% of the electromagnetic launcher
\begin{figure}
\centering
     \begin{subfigure}[b]{0.45\textwidth}
         \centering
       \includegraphics[width=1\linewidth]{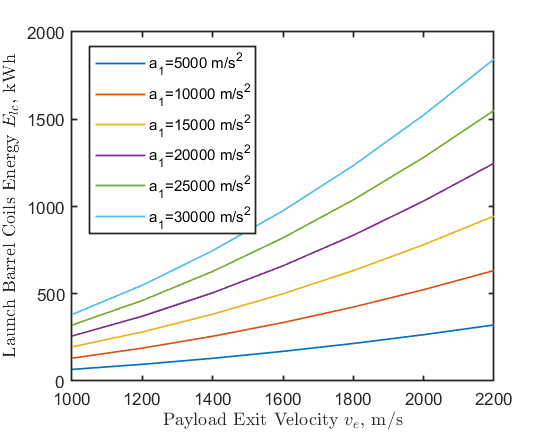}
        \caption{Comparison of $E_{lc}(v_e)$ on different launch acceleration $a_1$. }\label{fig:E_{lc}}  
\end{subfigure}
     \begin{subfigure}[b]{0.45\textwidth}
         \centering
  \includegraphics[width=1\linewidth]{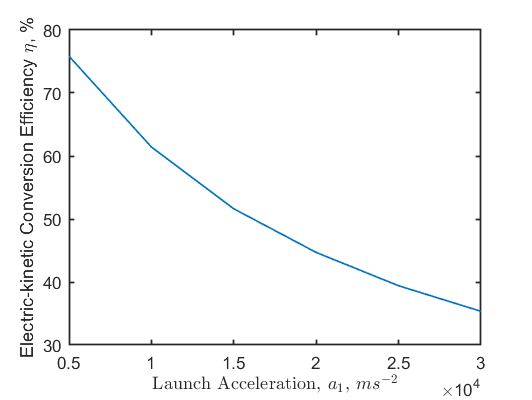}
        \caption{The Electric-kinetic conversion efficiency $\eta(a_1)$.}\label{fig:eta}      \end{subfigure}
        \caption{The launch barrel coils energy and the electric-kinetic conversion efficiency of the launch system.}
\end{figure}

This section shows how the payload acceleration $a_1$ and exit velocity $v_e$ affect the launch barrel coils energy consumption $E_{lc}$. We calculated the $E_{lc}$ via Eq. (\ref{E_lc}) on 42 cases. As shown in Figure \ref{fig:E_{lc}}, the launch barrel coils energy consumption increases as the payload exit velocity increases. Given an exit velocity, the larger payload acceleration requires more launch barrel coils energy consumption. While the electric-kinetic energy conversion $\eta$ of this launcher is defined in Eq. (\ref{eta}) \cite{engel2006efficiency}. As shown in Figure \ref{fig:eta}, $\eta$ decreases as the payload acceleration $a_1$ increases.
 \begin{align}
\label{eta}
    \eta = \frac{E_{ke}}{E_t},
\end{align}
where $E_{ke}$ is the payload final kinetic energy, $E_{lc}$ is the launch barrel coils energy consumption. The electric-kinetic conversion efficiency $\eta$ is defined as the ratio of the output energy and the total input energy. The output energy is obvious, the payload kinetic energy at the exit of the launcher. The total input energy includes the energy consumption in the coils to acceleration the payload, the contact energy losses, the resistance energy losses, and the converted kinetic energy. As we adopt the superconducting magnetic suspension launcher, the contact friction losses and the resistance energy losses are neglected. Therefore, the total input energy in the denominator is the sum of the kinetic energy of the payload and the energy consumption in the coils.

This electromagnetic launcher is more attractive than the propellent-based launcher in terms of the energy budget. For this design, it costs 1\$ to storage 360 J energy in an ultra-capacitor (UC) system, according to \cite{Battery}. 28.6 million USD is the budget for a 10,296 MJ energy storage system (the unit price and cycle life of SC are \$10,000/kWh and 30,000 hrs, respectively). This energy storage system can support up to 176 times launches because its cycle life is up to 30,000 hours (170 hours per launch). In this paper, the UC energy storage system costs about 0.16 million USD per launch (28.6/176=0.1625) to send a 1500 kg payload to the lunar orbit. It's about \$108/kg energy cost per kilogram of payload. While, the Apollo-Lunar-Module ascend stage can launch a 2347 kg payload to the lunar orbit and can take 2353 kg Aerozine 50 fule, which costs about 0.28 million USD. It's about \$122/kg energy cost per kilogram of payload. The fuel/energy cost of the Apollo-Lunar-Module ascend stage is a little higher than that of the electromagnetic launcher in the paper. However, the lunar surface's energy accessibility advantage is also evident for this design because solar radiation energy is more obtainable than rocket propellent chemical energy.  Furthermore, according to paper \cite{roy2019investigations}, it is assumed that the unit price and cycle life of SC are \$2,500/kWh and 500,000, respectively. The energy cost per kilogram payload of this design would further decrease to \$1.07/kg and allows a big reduction in the energy cost.
\par The energy consumption of the launch barrel coils (83 sections) is 1,852.2 kWh, so for a section of this launch tube, the energy request is about 22.316 (=1852.2/83) kWh. Choosing 350F ultra-capacitors (0.35 Wh each) for power supply, each launch tube section would need 63,759 (=22316/0.35) ultra-capacitors to power. These ultra-capacitors would connect in series. According to Eqs. (\ref{N_sb})-(\ref{I0}), and the value of the launch tube section coils inductance L=9480 H, the peak voltage, the current magnitude, and the peak power of each launch tube section are 172 kV, 131 A, and 22.4 MW, respectively. The total launch time is 0.08 s, this launch tube needs to turn on and off 166 (=83$\times$2 )times. So, the switch frequency requirement of this launcher is about 2,075 (=166/0.08) Hz.  For a high voltage direct current (HVDC) power transmission system, high voltage (HV) silicon carbide (SiC) power semiconductor device, such as an insulated-gate bipolar transistor (IGBT), have been developed and researched for high voltage (>35 kV) and fast switching speed (several kHz) applications \cite{ji2017overview,wang2016facts}. The voltage-source converter (VSC) is a promising evolving technology for power control driven by cost and efficiency.  It is worth mentioning that ABB has developed and installed the Estlink transmission system, which operates at 150 kV DC and is rated at 350 MW of active power via VSC-HVDC technology through IGBTs \cite{pan2008vsc}. For the power system parameters designed in this paper, it is reasonable for us to adopt similar VSC-HVDC technology through SiC IGBTs.  To get a good synchronization of the coils switching and the payload displacement, we need to use a switch-mode converter. The general switch-mode converter requires power processing circuitry, modulator (converts the control signal from the error amplifier to driving signals for the power switching devices), error amplifier compensator (compares the output to reference, generates an error signal, process error signal to create control signal and shapes frequency response to achieve stable control loop and dynamic response) and protection status monitoring communication system (voltage, current, temperature reporting, 8remote on/off, startup sequencing).

\section{Conclusions}
This paper designs and analyzes an electromagnetic tensegrity lunar launcher that can accelerate the lunar-derived commodities to a lunar exit velocity for space-depots utilization. The governing equations of design parameters are derived based on electromagnetic theories and tensegrity minimum mass principles. The design results are demonstrated by a case study. It is shown that to launch a 1,500 kg payload mass into lunar orbit at an exit velocity of 2,200 m/s with an acceleration of 3.0 $\times$ 10$^4$ m/s$^2$, the required launch height, propulsion force, and restoring force are 83.24 m, 46.5 MN, and 2,900 N in the presence of an assumed radial deflection of 5 mm. The required mass of the launch-barrel coil keeps at around 8.03$\times$ 10$^4$ kg when the launch acceleration is over 2.0 $\times$ 10$^4$ m/s$^2$. The launcher's total mass (sum of coil mass and structure mass) is 1.12 $\times$ 10$^5$ kg. In other words, the payload-mass-to-the-launcher-mass ratio is about 1.34\%. For a single launch, the total energy is estimated to be 2860.5 kWh, which will take a 250 m$^2$ solar panel approximately 170 hrs to get fully charged. The electromagnetic force calculation, the tensegrity launcher mass, and the launch energy are not limited to lunar applications. Other scenarios, such as high-speed trains, rocket launch systems, space transportation systems, are also applicable.

\section{Appendix}
\begin{figure}
        \centering
    \begin{subfigure}[b]{0.5\textwidth}
            \centering
    \includegraphics[width=0.8\linewidth]{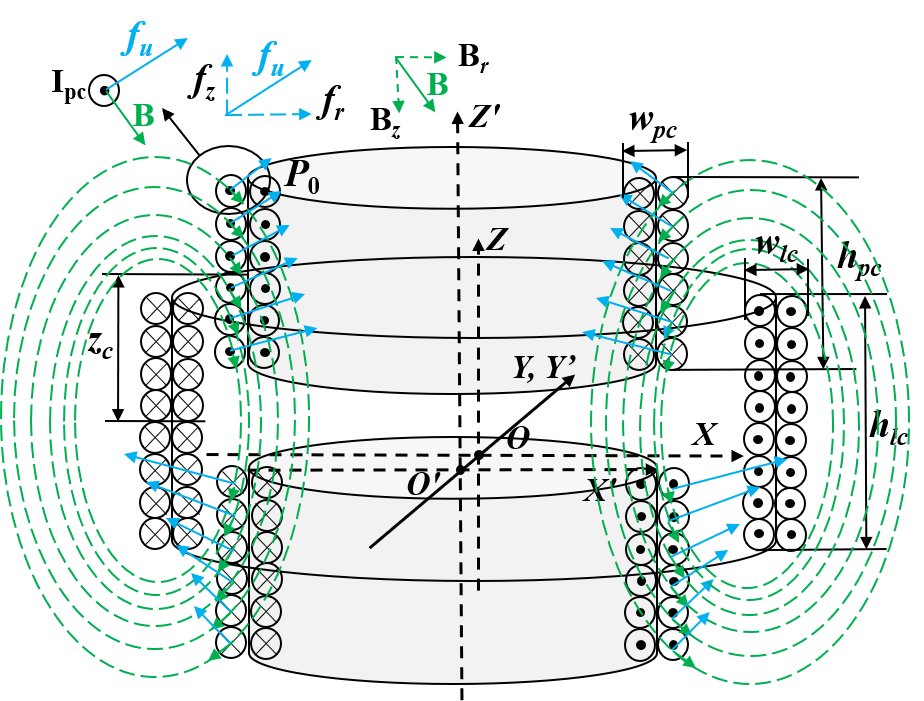}
    \caption{Configuration of the coils at one instant.}\label{fig:principle1} 
\end{subfigure}
\begin{subfigure}[b]{0.4\textwidth}
    \centering
    \includegraphics[width=0.8\linewidth]{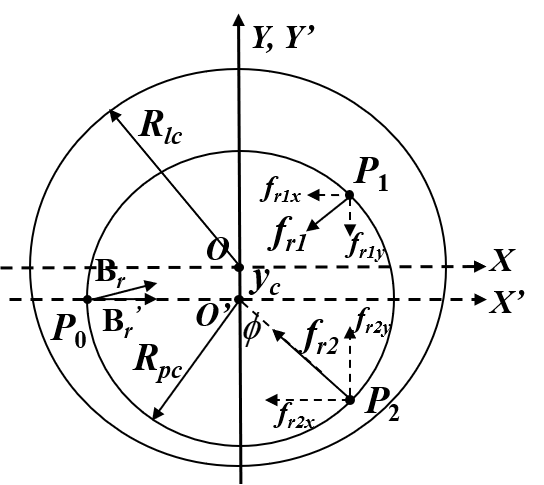}
    \caption{Restoring force acting on payload carrier coils.}\label{fig:principle2}
\end{subfigure}
\begin{subfigure}[b]{0.3\textwidth}
\centering
\includegraphics[width=1.1\linewidth]{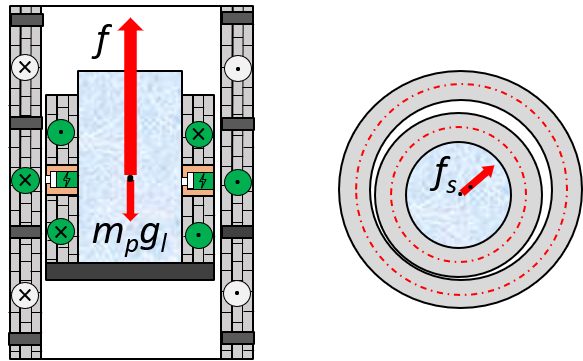}
\caption{Free body diagram of the payload illustrates the forces applied on the payload.}
\label{fig:FBD} 
\end{subfigure}
\caption{Schematic of the Electromagnetic propulsion principle.}
\end{figure}

\subsection{Electromagnetic propulsion principle and the calculation of the electromagnetic force}
This section provides the mathematical formulas to calculate the electromagnetic propulsion force $f$ and restoring force $f_s$ between the payload-carrier coils and the launch-barrel coils unit. The detailed derivations can be found in \cite{KiBong1996}.
A brief introduction on how electromagnetic forces are generated as a whole. As shown in Figures \ref{fig:principle1} and \ref{fig:principle2}, a magnetic field $B$ (vertical component $B_z$, radial component $B_r$, and horizontal projection $B_r'$) is generated by one current-carrying filament coil on the launch-barrel (outer coil). And then the electromagnetic force $f_u$ (vertical component $f_z$ and horizontal component $f_r$) is applied at point $P_0$ on one filament coil of the payload-carrier (inner coil) in the presence of $B$. Finally, we can obtain the total electromagnetic propulsion force $f$ and restoring force $f_s$ by integrating $f_z$ and $f_r$ at each point on each filament of the payload carrier produced by each filament coil on the launch-barrel coils unit. As shown in Figure  \ref{fig:FBD}, the free body diagram of the payload illustrates the forces applied on the payload, including the lunar gravity, the propulsion force $f$, and the restoring force $f_s$ exerted by the electromagnetic launcher.
Given the radius, height, width, current, and the total number of turns for the payload-carrier coils unit and the launch-barrel coils unit, the electromagnetic force ($f$ or $f_s$) can be obtained by multiplying a total electrical current with a unit electromagnetic force. The unit electromagnetic force is determined by the geometric relationships between the two coils and can be calculated via a filament coil method. The calculation includes the segmentation of the two coils, the filament mutual force, and their mutual forces' summation.\par 

1. Segmentation of the two coils: the width $w_{pc}$ of the payload-carrier coils unit is divided into $(2\lambda+1)$ segments, and the height $h_{pc}$ into $(2q+1)$ segments. Similarly, the launch-barrel coils unit has $(2\Lambda+1)\times(2Q+1)$ filaments. Note that we use upper and lower cases to represent the number of segments for the launch barrel coils and the payload carrier coils, respectively.\par 

2. The filament mutual forces ( propulsion force $f_{zu}$ and restoring force $f_{su}$) between the $(k^{th},l^{th})$ filament on the payload-carrier coils unit and the $(i^{th},j^{th})$ filament on the launch-barrel coils unit are:
\begin{align} \nonumber
     f_{zu(i,j,k,l)} & = \frac{\mu_0 R_{pc} z_c}{2\pi}\int_{-\pi/2}^{\pi/2}[\frac{1}{r_{pc,l}^{1}}(\frac{R_{lc,j}^2+(r_{pc,l}^{1})^2+z_{i,k}^2}{(R_{lc,j}-r_{pc,l}^{1})^2+z_{i,k}^2} \\ \nonumber & \text{E}(k_1^2)-\text{K}(k_1^2))\frac{\text{cos}(\phi-\gamma_1)}{\sqrt{(R_{lc,j}+r_{pc,l}^{1})^2+z_{i,k}^2}}\\ \nonumber
    &+\frac{1}{r_{pc,l}^{2}}(\frac{R_{lc,j}^2+(r_{pc,l}^{2})^2+z_{i,k}^2}{(R_{lc,j}-r_{pc,l}^{2})^2+z_{i,k}^2}\text{E}(k_2^2) \\&  -\text{K}(k_2^2))\frac{\text{cos}(\gamma_2-\phi)}{\sqrt{(R_{lc,j}+r_{pc,l}^{2})^2+z_{i,k}^2}}]d\phi,
\end{align}
\begin{align} \nonumber
 f_{su(i,j,k,l)} &= \frac{\mu_0 R_{pc}}{2\pi} \int_{-\pi/2}^{\pi/2}[(\frac{R_{lc,j}^2-(r_{pc,l}^{1})^2-z_{i,k}^2}{(R_{lc,j}-r_{pc,l}^{1})^2+z_{i,k}^2}\text{E}(k_1^2) \\ \nonumber & +\text{K}(k_1^2))\frac{\text{cos}\phi}{\sqrt{(R_{lc,j}+r_{pc,l}^{1})^2+z_{i,k}^2}}\\ \nonumber
& -(\frac{R_{lc,j}^2-(r_{pc,l}^{2})^2-z_{i,k}^2}{(R_{lc,j}-r_{pc,l}^{2})^2+z_{i,k}^2}\text{E}(k_2^2) +\text{K}(k_2^2)) \\ & \frac{\text{cos}\phi}{\sqrt{(R_{lc,j}+r_{pc,l}^{2})^2+z_{i,k}^2}}]d\phi,
 \end{align}
where 
\begin{align} 
z_{i,k} & = z_c-i\frac{h_{lc}}{2Q+1}+k\frac{h_{pc}}{2q+1}, \\ R_{lc,j} & =R_{lc}+j\frac{w_{lc}}{2\lambda+1},~
R_{pc,l} =R_{pc}+l\frac{w_{pc}}{2\lambda+1},\\
r_{pc,l}^{1} & = \sqrt{(R_{pc,l}\text{cos}\phi+y_c)^2+(R_{pc,l}\text{sin}\phi)^2}, ~ \\
r_{pc,l}^{2} & =\sqrt{(R_{pc,l}\text{cos}\phi-y_c)^2+(R_{pc,l}\text{sin}\phi)^2},\\ 
k_1^2 & =\frac{4R_{lc,j}r_{pc,l}^{1}}{(R_{lc,j}+r_{pc,l}^{1})^2+z_{i,k}^2}, \\
k_2^2 & =\frac{4R_{lc,j}r_{pc,l}^{2}}{(R_{lc,j}+r_{pc,l}^{2})^2+z_{i,k}^2}, \\ 
\gamma_1  &=  \frac{R_{pc,l}\text{sin}\phi}{R_{pc,l}\text{cos}\phi+y_c}, ~\gamma_2 = \frac{R_{pc,l}\text{sin}\phi}{R_{pc,l}\text{cos}\phi-y_c},
\end{align} 
where $\text{K}$, $\text{E}$ are the complete elliptic integrals of the first and second kind. $f_{zu}$ and $f_{su}$ are filament propulsion force and filament restoring force. $R_{lc,j}$ is the radius of the $j^{th}$ layer on the launch-barrel coils, $r_{pc,l}$ is the radius of the $l^{th}$ layer on the payload-carrier coils. $z_{i,k}$ is the height between the $i^{th}$ layer on the launch-barrel coils and the $k^{th}$ layer on the payload-carrier coils. $\phi$ is the integral dummy variable to calculate electromagnetic force on each coil filament of an unit payload-carrier coils. $\gamma_1$ and $\gamma_2$ are two angles to calculate magnetic field projections. $R_{pc}$ and $R_{lc}$ are the payload-carrier radius and the launch-barrel radius. $w_{pc}$ and $w_{lc}$ are the width of the payload-carrier coils and the launch-barrel coils. $h_{pc}$ and $h_{lc}$ are the height of the payload-carrier coils unit and the height of the launch-barrel coils unit. $z_c$ is a height-center distance between an unit payload-carrier coils and an unit launch-barrel coils. $y_c$ is a misalignment between payload-carrier coils and launch-barrel coils. $\mu_0$ is the space permeability 1.257 $\times$ 10$^{-6}$ H/m.
\par
3. The unit electromagnetic forces (the unit propulsion force $G_f$ and the unit restoring force $G_s$) are calculated by summation of the filament mutual forces:
\begin{align}
   \nonumber
& G_f(y_c,z_c,R_{pc},R_{lc},h_{pc},h_{lc},w_{pc},w_{lc})= \\ \label{eqn:G_f}  
& \sum_{i=-Q}^{Q}\sum_{j=-\lambda}^{\Lambda}\sum_{k=-q}^{q}\sum_{l=-\lambda}^{\lambda}\frac{f_{zu(i,j,k,l)}}{N_e},\\ \nonumber
& G_s(y_c,z_c,R_{pc},R_{lc},h_{pc},h_{lc},w_{pc},w_{lc}) = \\
& \sum_{i=-Q}^{Q}\sum_{j=-\Lambda}^{\Lambda}\sum_{k=-q}^{q}\sum_{l=-\lambda}^{\lambda}\frac{f_{su(i,j,k,l)}}{N_e}, \\
& N_e = (2Q+1)(2\Lambda+1)(2q+1)(2\lambda+1),
\end{align}
where $G_s$ and $G_f$ are the unit restoring force function and the unit propulsion force function. $N_e$ is the total number of iterations to calculate electromagnetic forces.

4. The total electromagnetic forces can be obtained by multiplying the unit electromagnetic forces by the total electrical currents:
\begin{align}
\nonumber
f  = &  2N_{pc} N_{lc} I_{pc} I_{lc}G_f(y_c,z_c,R_{pc},R_{lc},\\ \label{propulsionforce} & h_{pc},h_{lc},w_{pc},w_{lc}),\\ \nonumber
f_s  = & 2N_{pc} N_{lc} I_{pc} I_{lc}G_s(y_c,z_c,R_{pc},R_{lc},\\ &h_{pc},h_{lc},w_{pc},w_{lc}),
\label{restoringforce}
\end{align}
where $I_{pc}$ and $I_{lc}$ are the current in each turn of coils of the payload-carrier coils and the launch-barrel coils. $N_{pc}$ and $N_{lc}$ are the number of turns of an unit payload-carrier coils and an unit launch-barrel coils.

\section{Acknowledge}
We thank Dr. Bo Liu, Sr. research engineer in power electronics at Raytheon Technologies Research Center, for suggestions with high voltage fast switching power control strategies that greatly improved the manuscript.

\bibliographystyle{cas-model2-names}

\bibliography{cas-refs}
% \bibliography{cas-refs.bib}

\end{sloppypar}
\end{document}